\documentclass[natbib]{mn2e}

\usepackage{natbib}
\usepackage{graphicx}
\usepackage{multirow}
\usepackage{rotating}
\usepackage{txfonts}

\title[Potential TTV detection biases]{Potential Biases in the Detection of Planetary Systems with Large Transit Timing Variations}
\author[E. Garc\'ia-Melendo and M. L\'opez-Morales]{E. Garc\'ia-Melendo$^{1,2}$\thanks{E-mail:
egarcia@foed.org, mlopez@ieec.uab.es} and M. L\'opez-Morales$^{2,3}$\\
$^{1}$Fundaci\'o Privada Observatori Esteve Duran, 08553 Seva, Spain\\
$^{2}$Institut de Ci\`encies de L'Espai, Campus UAB, Fac. Ci\`encies, Torre C5 p2, 08193 Bellaterra, Barcelona, Spain\\
$^{3}$Visiting Investigator. Carnegie Institution of Washington, Department of Terrestrial Magnetism, 5241 Broad \\   Branch Road NW, Washington, DC 20015, USA}
\begin{document}

\date{ }

\pagerange{\pageref{firstpage}--\pageref{lastpage}} \pubyear{2011}

\maketitle

\label{firstpage}

\begin{abstract}
The Transit Timing Variations (TTVs) technique provides a powerful tool to detect additional planets in transiting exoplanetary systems. In this paper we show how transiting planets with significant TTVs can be systematically missed, or cataloged as false positives, by current transit search algorithms, unless they are in multi-transit systems. If the period of the TTVs, $P_{TTV}$, is longer than the time baseline of the observations and its amplitude, $A_{TTV}$, is larger than the timing precision limit of the data, transiting planet candidates are still detected, but with incorrect ephemerides. Therefore, they will be discarded during follow-up. When $P_{TTV}$ is shorter than the time baseline of the observations and $A_{TTV}$ is sufficiently large, constant period search algorithms find an average period for the system, which results in altered transit durations and depths in the folded light curves. Those candidates can get subsequently discarded as eclipsing binaries, grazing eclipses, or blends. Also, for large enough $A_{TTVs}$, the transits can get fully occulted by the photometric dispersion of the light curves. These detection biases could explain the observed statistical differences between the frequency of multiple systems among planets detected via other techniques and those detected via transits. We suggest that new transit search algorithms allowing for non-constant period planets should be implemented.
\end{abstract}


\section{Introduction}
In a transiting exo-planetary system, the presence of other planets can be inferred by the effect of their gravitational pull on the transiting planet. That effect can be detected as periodic transit timing variations (TTVs), with amplitudes potentially as large as several tens of minutes \citep[e.g][]{Miralda2002, Agol2005, Holman2005, Heyl2007, Veras2010}.

These theoretical predictions have been confirmed by two recent results from the Kepler satellite mission \citep{Borucki2004}. In the first of those results, the system Kepler- 9, composed of two transiting Saturn-size planets, presents TTVs with average amplitudes of 4 and 39 minutes 
per orbit \citep{Holman2010}. The other reported system, Kepler-11, composed of five $2.3 - 13.5 M_{Earth}$ transiting planets within 0.25AU from the star, and a sixth more massive planet in an outer orbit, shows TTVs amplitudes of up to 30 minutes \citep{Lissauer2011}. \citet{Steffen2010} have also published five Kepler systems with more than one transiting object, although those have not been confirmed as planets.

It is surprising that TTVs have been first detected in multiple transit planet systems, i.e. systems where several planets are co-planar within a few degrees, since those are geometrically less likely than non-coplanar multiple planet systems where only one planet transits. However, no definitive detection of single-transiting planets TTVs has been reported. 
%until just recently \citep[see HAT-P-13b's TTV detection by ][]{Nascimbeni2011,Pal2011}. 
Some authors have reported hints of possible TTVs, for WASP-3b \citep{Maciejewski2010}, WASP-10b \citep{Maciejewski2011}, WASP-5b \citep{Fukui2011}, and most recently HAT-P-13b \citep{Nascimbeni2011,Pal2011}, but none of those claims  has yet been confirmed. \citet{Ford2011} also recently reported over 60 TTVs candidate systems in the Kepler mission data.

The apparent scarcity of single-transit multi-planet systems was also noticed by \citet{Fabrycky2009}, who showed how one third of the non-transiting planets discovered via the radial velocity technique, and with orbital periods similar to the transiting planets known at the time, had companions versus no detected companions for any of the transiting planets. Two single-transit planetary systems with a second companion have been detected since then, CoRoT- 7b \citep{Leger2009, Queloz2009} and HAT-P-13b \citep{Bakos2009}, but the statistics still disagree.

In this work we explore potential observational and transit search systematics that could introduce single-transit multi-planet systems detection biases. Section 2 summarizes how the transit search algorithms used by successful surveys work and their potential detection biases. Section 3 describes our model simulation for a generic single-transiting system with TTVs and its main observational signatures. Section 4 describes the model simulation results. We discuss our findings in section 5.

\section{Transit Search Surveys and Algorithms}\label{sec:surveys}

%% Figure 1
\begin{figure*}
%{\centering
  \includegraphics[width=170mm]{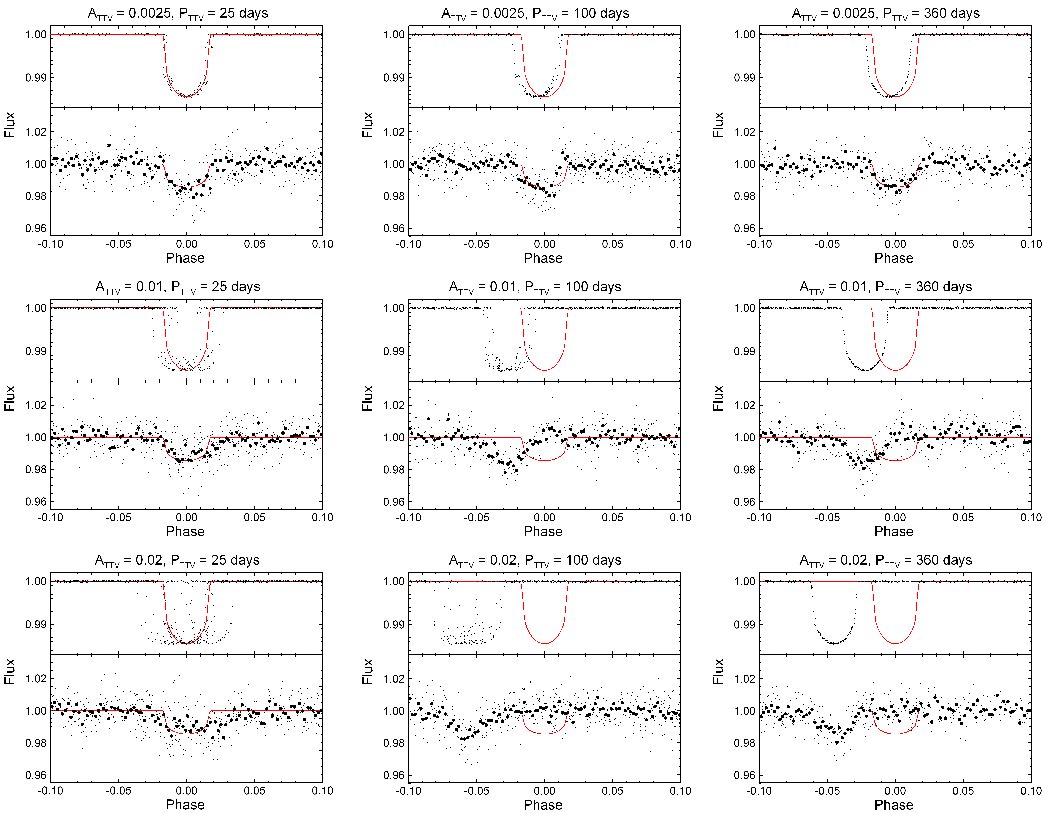}
\caption{Effect of TTVs for 90-day observing blocks, when $P_{TTV}$ = 25, 100 and 360 days (columns left to right) and $A_{TTV}$ = 0.0025, 0.01 and 0.02 (rows top to bottom). In each panel, the top curves are model data for $\sigma_{ph} = 10^{-4}$ (Kepler case), and the bottom curves are for  $\sigma_{ph} = 8 \times 10^{-3}$ (ground-based case). All curves are folded over the best-fit periods. Solid circles show ground-based model data binned over 0.002 phase intervals.}
 \label{models1}
%}
\end{figure*}

Since the discovery of the first transiting planet HD209458b a decade ago \citep{Charbonneau2000, Henry2000}, several surveys have successfully detected many other transiting systems. All those surveys combined, i.e. the ground- based projects OGLE \citep{Udalski1992}, HAT \citep{Bakos2002}, TrES \citep{Dunham2004, Alonso2004}, XO \citep{McCullough2005}, SuperWASP \citep{Pollacco2006}, and the recent space missions CoRoT \citep{Auvergne2009} and Kepler \citep{Borucki2004}, have already discovered over 100 transiting planets. In addition, the Kepler mission has recently announced 1235 transit candidates \citep{Borucki2011b}, of which 90$\%$ are expected to be true exoplanets \citep{Morton2011}.

Transit planet candidates in each of those surveys are identified via custom-made search algorithms adapted to the specific characteristics of each survey (i.e. noise systematics, sampling rate, etc ...). All the algorithms follow common approaches, which either look at the unfolded light curves in search for transit-like signals, or fold the light curves over trial periods until a transit-like signal is found. Because of the typically sparse sampling of single orbit light curves, the second approach is the one ground-based surveys generally use. All the above ground-based surveys use the Box-fitting Least Square (BLS) algorithm \citep{Kovacs2002} or some modified version of it. The BLS algorithm consists on folding the light curves over a series of periods, and then fitting the folded curves with narrow box-shaped functions of different depths and widths until finding the best least-square fit. OGLE, TrES and HAT use this approach. XO uses a modified form of the BLS algorithm which includes an adaptive-mesh search routine to increase the speed at which optimal periods are identified \citep{McCullough2005}. SuperWASP also uses a modified form of BLS which selects the strongest peaks of the BLS periodogram via Newton-Raphson refinement \citep{Protopapas2005}. They also refine the box-shape function approach by 1) estimating the planetary radius using the physical properties of the star and analytical approximations between the transit depth and the stellar and planetary radii, and 2) inspecting the folded light curves for photometric ellipsoidal variations characteristic of stellar binaries \citep{Collier2006}.

The CoRoT transit search algorithms have been developed by different participating teams and applied independently to the data. The algorithms, described in detail in \citet{Moutou2005}, \citet{Regulo2007}, \citet{Carpano2008} and \citet{Renner2008}, follow approaches such as correlations with sliding transit templates, box-shaped signal searches, wavelet transformations or BLS. Some of the transit search routines in these algorithms work with folded data, while others use the unfolded light curves. However, all the algorithms, except for the ones by \citet{Renner2008} and \citet{Regulo2007}, limit their analysis to strictly periodic signals. The \citet{Renner2008} algorithm includes some tolerance for period variations by allowing period error margins of 0.1 days. The \citet{Regulo2007} algorithm, based on wavelet transformations, can detect period variations of about 2$\%$, depending on the precision of the data (R\'egulo, priv. comm.). Both algorithms use unfolded light curves.

Finally, Kepler uses a custom-made transiting planet search (TPS) pipeline, which starts by applying a wavelet-based, adaptive matched filter to minimize the light curves noise, followed by a transit waveform (i.e. box-shape-like function) to search for individual transit events in the unfolded data. Those events are then folded over a wide range of periods until the period that best matches the full set of events is found \citep[see][]{Jenkins2002,Jenkins2010a,Jenkins2010b,Borucki2011a}. TPS, which is optimized to detect Earth-size planets, folds the transit candidate events at strict periodicities, not allowing for very large TTVs. However, because of the continued monitoring of targets, the pipeline does flag systems with significant TTVs (e.g. Kepler-9, Kepler-11).

\section{Model Simulations}\label{sec:models}

We have conducted model simulations to assess the effectiveness of current transit search algorithms for the case of transiting planets with significant TTVs. Our model system consists of a Jupiter-size planet orbiting a Sun-like star, with an orbital period $P_p$ = 3.6235 days and impact parameter $b$ = 0. The model transit, generated using $jktebop$ \citep{Southworth2004a, Southworth2004b}, and a linear limb darkening of 0.61, has a depth of 1.3$\%$, easily detectable by both ground and space surveys. Notice that we use as example case
a model with parameters typical of known transiting hot Jupiters for the purpose of illustrating the
potential TTV detection biases. The effect is still the same for planets with smaller radius or larger transit impact parameter, except for the fact that transit detection limits will depend more strongly on the noise levels of the data (i.e. shallower transits get diluted into the noise at increasingly lower values of $A_{TTV}$, see \S 4).

Using that model we simulate a series of transits adopting as variable parameters the length of the observing blocks, the observing cadence, the photometric dispersion of the light curves, $\sigma_{ph}$, and the amplitude and period of the TTVs, $A_{TTV}$ and $P_{TTV}$ . In our simulations $A_{TTV}$ is defined as a fraction of the orbital period of the planet, or equivalently, a fraction of the orbital phase. For the TTVs we assume a simple periodic model of the form

\begin{equation}
T_{mid} = T_0 + nP_p +A_{TTV}P_p \sin \left ( \frac{2\pi}{P_{TTV}} n P_p \right ), 
\end{equation}

\noindent where $P_{TTV}$ is given in days, $n$ is the number of elapsed orbits with respect to an arbitrary initial transit epoch, $T_0$, and $T_{mid}$ is the mid-transit time of each transit.

For the simulations we adopt values of $A_{TTV}$ between 0.0025 and 0.02,	and values of $P_{TTV}$ between 25 and 360 days. In addition, to resemble as close to real observing conditions as possible, we perform the simulations with parameters typical of both ground-based and space-borne surveys. In the case of ground-based surveys we simulate two scenarios: the first one adopts an observing block length of about 3 months (90 days), a photometric dispersion $\sigma_{ph} = 8 \times 10^{-3}$, and an observing cadence of 8 minutes, which approximate the average values of those parameters for the ground-based surveys listed in section 2, over one observing season (year). In the second scenario we generate three observing blocks of 90 days each over a three-year period 
%(i.e. three observing seasons for a given target)
 to simulate long-term ground-based observations spanning over more than one observing season. In the simulations we also include the effect of day-night cycles, and a 70$\%$ probability of clear nights to estimate the total number of observable transits. To simulate the conditions of a space-borne survey, we adopt parameters similar to the Kepler mission, i.e. a photometric dispersion $\sigma_{ph} = 10^{-4}$, a 30 minute cadence, and two different observing block lengths: one of 90 days, to compare with similar ground-based data, and another one with continuous observations over 3 years, which is the nominal lifetime of the mission \citep{Borucki2004}.
 
 Each synthetic light curve is then folded over a wide range of trial periods, in a manner similar to the BLS algorithms, until the best period is found. That best period is designated as the one that minimizes the transit duration in the folded light curves, or equivalently, period  that minimizes the transit's photometric dispersion. To interpret our results we use the raw space-borne model data, and the ground-based model data averaged over 0.002 phase bins.

%% Figure 2
\begin{figure}
{\centering
  \includegraphics[width=0.48 \textwidth,angle=0]{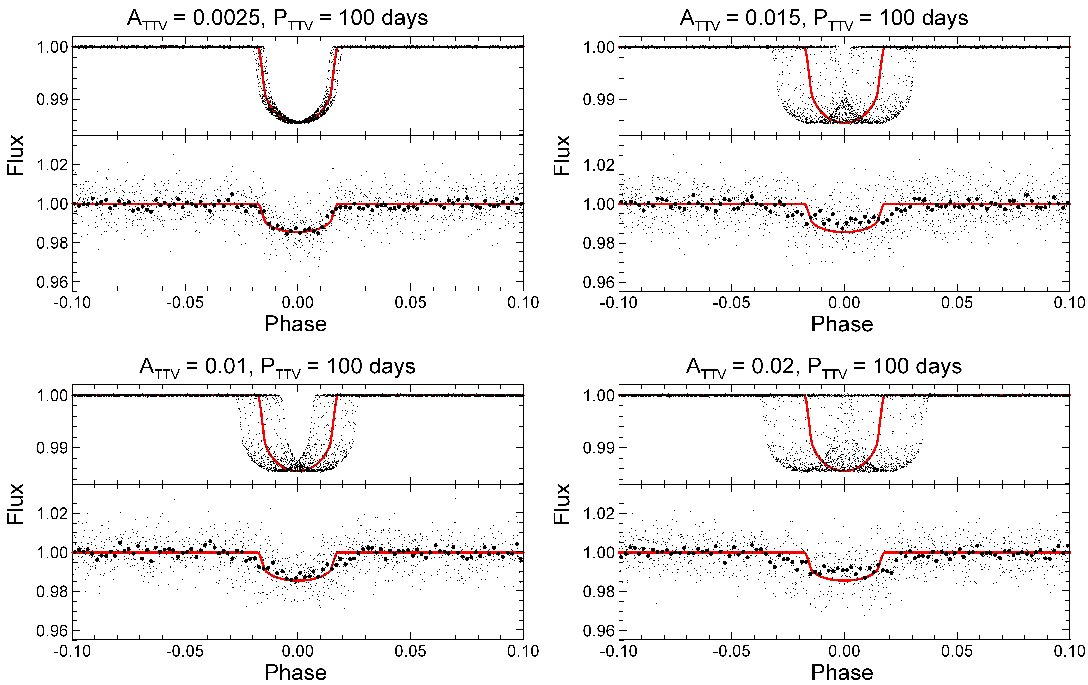}}
\caption[]{Same as in figure 1, but for the 3-year observing blocks. 
All plots correspond to $P_{TTV}$ = 100 days. The amplitudes of the 
TTVs represented are $A_{TTV}$ = 0.0025, 0.01, 0.015 and 0.02. \label{models2}
}
\end{figure}

\section{Simulation Results}\label{sec:results}

The outcome of the simulations is illustrated in Figures 1 and 2. Figure 1 shows the results for the 90-day observing blocks, both for the ground-based and Kepler survey cases. Figure 2 shows the results for the 3-year observing blocks (3 seasons for ground-based surveys and continuous observations for Kepler). In all cases, a BLS-type algorithm should detect transit-like dips for $A_{TTV} \lesssim 0.02$. However, the characteristics of the curves change with the values of $A_{TTV} $ and $P_{obs}$, i.e. the length of the observing blocks. 

Figure 1 reveals two distinct cases. When $P_{TTV} < P_{obs}$ (left column in figure), the full amplitude of the TTV is observed and any best-fit algorithm should fold the data about the correct orbital period of the planet, $P_p$. However, the transits widen and get shallower as $A_{TTV}$ increases. While the effect of these TTVs will be discernible in Kepler-like data, those transiting planet candidates will be most likely labeled as false positives given the lower light curve data quality of a typical ground-based survey (signals that wide and shallow will be mistaken for grazing binary eclipses).  Further tests show that for $A_{TTV} \gg 0.02$ the transits get completely diluted in typical ground-based data. When $P_{TTV} > P_{obs}$ (right column in figure), the algorithms find a perfect transit shape, but with the data folded over an erroneous period

\begin{equation}
P_{p}^{'}  = P_p  \left ( 1+ 2 \pi A_{TTV} \frac{P_p}{P_{TTV}} \right ), 
\end{equation}

\noindent which results from eq. 1 when $2 \pi \frac{nP_p}{P_{TTV}} \ll 1$. This period offset also affects the $T_0$ zero epoch value, and therefore the resulting transit ephemerides equation will be incorrect. Any follow-up observation will therefore miss the transit event and the candidate will be discarded as a spurious detection.

For intermediate cases, where $P_{TTV} \sim P_{obs}$ (middle column in figure), a best-fit folded light curve simultaneously yields a wrong ephemeris and shape-deformed minima, which may again elude correct detection. 

Figure 2 illustrates a situation similar to the left-side column of figure 1 (i.e. $P_{TTV} < P_{obs}$), but for the 3-year observing blocks. This figure shows more clearly how the depth and width of the transits get altered as a function of $A_{TTV}$. In the case of $A_{TTV}$ = 0.01, for example, the folded ground-based light curve resembles a "V" shape, typical of binary star false positives.

%% Figure 3
\begin{figure}
{\centering
  \includegraphics[width=0.48 \textwidth]{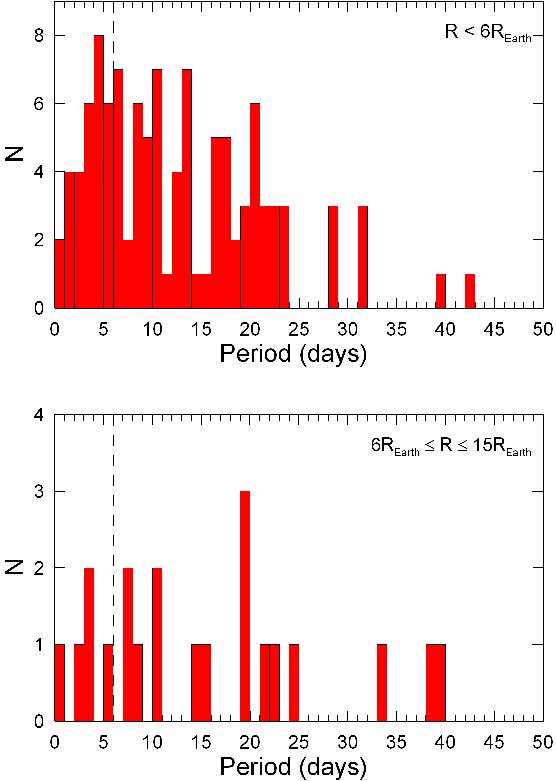}}
\caption[]{Histograms depicting the frequency of planets with TTVs in the \citet{Ford2011} sample as a function of orbital period. The top panel contains all planets smaller than about Neptune size ($R_p < 6 R_{Earth}$); the bottom panel contains all Saturn-Jupiter size planets ($6R_{Earth} < R_p < 15 R_{Earth}$. The vertical dashed line around P = 6 days indicates the, by definition, period limit of hot Jupiters. Most of the TTV detections for close-in planets correspond to the smaller planets, but there are also several massive close-in planets in the sample, in contrast with theoretical expectations.
\label{RvsP}
}
\end{figure}

%% Figure 4
\begin{figure}
{\centering
 \includegraphics[width=0.48 \textwidth,angle=0]{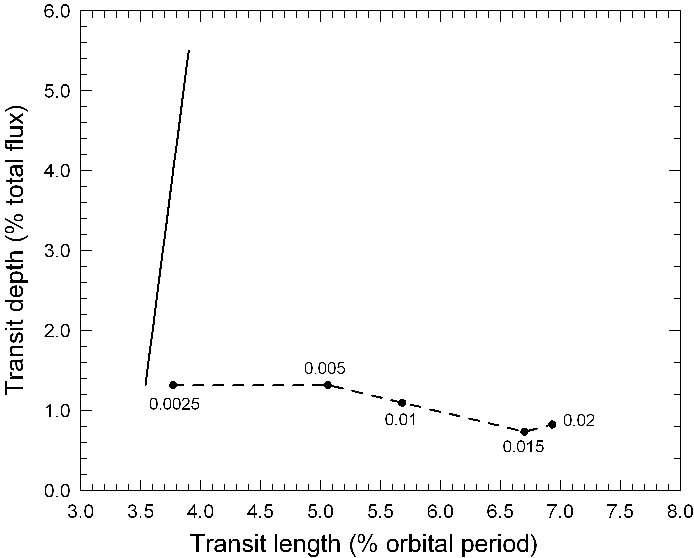}}
\caption[]{Representation of the transit depth versus transit length for hypothetical planets with radius between 1 and 2 $R_{jup}$ (solid line). The dashed line shows the same relation derived from the result of our simulations for $A_{TTV}$ = 0.0025, 0.005, 0.01, 0.015 and 0.02, a 3-year observing block length, and $P_{TTV}$ 100 days.\label{DvsW}
}
\end{figure}

\section{Discussion}\label{sec:discussion}

\citet{Terquem2007} and \citet{LoCurto2010} have suggested that the absence of multiple planet systems (and therefore TTVs) among transiting planets discovered by surveys (in their majority hot Jupiters discovered from the ground until Kepler was launched) could be due to migration mechanisms. Massive short period planets would have moved inwards disturbing the orbits of all other planets in their way. However, the recent detection with Kepler of over 130 candidate systems showing some evidence of TTVs challenges those hypotheses \citep{Ford2011}. Figure 3 represents the frequency of TTV candidates reported by \citet{Ford2011} in two separate histograms; one for planets with $R_p < 6 R_{Earth}$, i.e. Neptune-size and smaller planets, and the other for planets with $6 R_{Earth} < R_p < 15 R_{Earth}$, i.e. gas giants. Given photometric precision and observing cycle limitations, gas giants with orbital periods of six days or less are the planets most likely to be detected from the ground. The bottom diagram in the figure shows five TTV candidates that match those criteria. 
%In addition, the recent result by \citet{Nascimbeni2011} reveals TTVs for the hot Jupiter HAT-P-13b with an amplitude of $\sim 20$ minutes, i. e. $A_{TTV} \sim 0.005$. 
Assuming all of the \citet{Ford2011} candidates truly are planets with TTV variations, and comparing those numbers with the total number of hot Jupiters candidates with $P < 6$ days in the current Kepler mission's candidate list ($\sim 65$ planets), 1 out of each 15 transiting hot Jupiters should  show TTVs. Based on this estimate and given the number of hot Jupiter transiting planets currently confirmed, about 7 systems with significant TTVs should have been already discovered from the ground. Doing a similar estimation with the number of multiple planet systems found to date via the radial velocity technique (source http://exoplanet.eu), we find that 14 have planets with orbital periods $P < 6$ days and $Msini > 0.5 M_J$. Of those 2 are Jupiter-mass planets (assuming $Msini > 0.5 M_J$), which gives a fraction of close-in massive planets in multiple systems of about 1/7, although not all will present significant TTVs . Assuming also that a fraction of the \citet{Ford2011} TTV candidates will be false detections, we conclude that the true number of hot Jupiters showing TTVs falls somewhere below 1/15, and those systems are most likely being missed by current planet search algorithms. 

%The  \citet{Nascimbeni2011} result indicates that at least $\sim 1/100$ hot Jupiters presents TTVs, Therefore, the true number must fall somewhere in between, 

Our simulations reveal two effects that may explain why transiting planets with significant TTVs elude detection. One is finding the wrong transit ephemeris when $P_{TTV} > P_{obs}$. This effect does not preclude those systems from being labeled as transit candidates, but if a dubious candidate around, for instance, a relatively dim star is detected by a small telescope with $\sigma_{ph} \gtrsim 0.01$, and cannot be confirmed by observing it at a higher photometric precision because the ephemerides are incorrect, the candidate will be discarded before spectroscopic confirmation follow-up.

The other effect, i.e. wider, shallower and "V"-shaped transits, may explain why planets with $P_{TTV} < P_{obs}$ and significantly large $A_{TTV}$ are lost. Figure 4 shows how, for our simulated 1$R_{jup}$ planet, the depth and length of the transit changes in the folded light curve when $A_{TTV}$ varies from 0.0025 to 0.02. The almost vertical line in the figure indicates, for comparison, how the transit length and depth of a real planet would vary when its radius increased from 1 to 2 $R_{jup}$. Clearly, a modified BLS algorithm , or visual inspection, will find the depth and length of the candidate to be incompatible with the expected parameters of a real planet, and the system will be rejected as a grazing eclipsing binary. 
In the case of very low light curve scatter and continuous space-borne observations, like Kepler's, TTVs will be more difficult to miss, as show Figures 1 and 2. However, faint targets will encounter the same problems as ground-based observations.  Algorithms like the ones developed by \citet{Regulo2007} and \citet{Renner2008} should have been able to detect TTVs in the CoRot data, but the limited monitoring per field \citep[$\sim$ 150 days,][]{Auvergne2009}, and strong data systematics may have prevented such detections. 

We conclude that, if our hypotheses are correct, the implementation of new transit search algorithms optimized for the detection of TTVs, and their application to existing ground-based transit-search databases, should lead to the detection of new single-transit systems with significant TTVs. 

\section*{Acknowledgments}

We thank I. Ribas for useful discussions during the development of this work. We also thank J. Jenkins, D. Latham, C. R\'egulo and S. Renner for helpful clarifications about transit search algorithms of different surveys.

%\bibliographystyle{mn2e}
%\bibliography{ref}

\end{document}